\documentclass[aps,twocolumn,showpacs,preprintnumbers,prl]{revtex4}
\usepackage{graphicx}

\newcommand{\R}{{\bf r}}
\newcommand{\RR}{{\bf r}'}
\newcommand{\RP}{{\bf r}''}

\newcommand{\be}{\begin{equation}}
\newcommand{\ee}{\end{equation}}
\newcommand{\bea}{\begin{eqnarray}}
\newcommand{\eea}{\end{eqnarray}}
\newcommand{\bean}{\begin{eqnarray*}}
\newcommand{\eean}{\end{eqnarray*}}

\begin{document}

\title{High-Level Correlated Approach to the Jellium Surface Energy,\\
                      Without Uniform-Electron-Gas Input}
\author{Lucian A. Constantin$^1$, J. M. Pitarke$^{2,3}$, J. F. Dobson$^4$, A. 
Garcia-Lekue$^1$, and John P. Perdew$^5$}
\affiliation{
$^1$Donostia International Physics Center (DIPC),\\
Manuel de Lardizabal Pasealekua, E-20018 Donostia, Basque Country\\
$^2$CIC nanoGUNE Consolider, Mikeletegi Pasealekua 56, E-2009 Donostia, Basque
Country\\
$^3$Materia Kondentsatuaren Fisika Saila, UPV/EHU, and Unidad F\'\i
sica Materiales
CSIC-UPV/EHU,\\
644 Posta kutxatila, E-48080 Bilbo, Basque Country\\
$^4$Nanoscale Science and Technology Centre, Griffith
University, Nathan, Queensland 4111, Australia,\\ $^5$Department of Physics and
Quantum Theory Group, Tulane University, New Orleans, LA 70118}

\date{\today}

\begin{abstract}
We resolve the long-standing controversy over the surface energy of
simple metals: Density functional methods that require uniform-electron-gas
input agree with each other at many levels of
sophistication, but not with high-level correlated calculations like
Fermi Hypernetted Chain and Diffusion Monte Carlo (DMC) that
predict the uniform-gas correlation energy.  Here we apply a very
high-level correlated approach, the inhomogeneous Singwi-Tosi-Land-Sj\"olander
(ISTLS) method, and find that the density functionals
are indeed reliable (because the surface energy is "bulk-like").
ISTLS values are close to recently-revised DMC values.  Our work
also vindicates the previously-disputed use of uniform-gas-based
nonlocal kernels in time-dependent density functional theory. 
\end{abstract}


\maketitle

Density-functional theory (DFT)~\cite{KS} provides ground-state electron
densities and energies (or, in its time-dependent version
(TDDFT)~\cite{tddft}, excitation energies) for atoms, molecules, and solids.
Because of its simple selfconsistent-field structure, DFT is used for
electronic-structure calculations almost exclusively in condensed matter
physics, and heavily in quantum chemistry. Exact in principle, the
theory requires in
practice approximations for the exchange-correlation (xc) energy (or for the
xc kernel) as a functional of the density.  All commonly-used nonempirical
approximations require input from the uniform electron gas, which is
transferred to inhomogeneous densities.  The reliability of these
approximations must be judged a posteriori, and there is a long-standing
puzzle related to their reliability for solid surface energies, with
implications for vacancies and clusters~\cite{PWE}. 

The surface energy $\sigma$ is the energy cost per unit area to split a bulk
solid along a plane. This is not only of technological importance but also a
classic and highly sensitive test case for theories of exchange and
correlation in many-electron systems. The simplest model is jellium, in which
a uniform positive background of density $n=3/4\pi r_s^3=k_F^3/3\pi ^3$
terminates sharply at a plane and is neutralized by electrons that penetrate
into the vacuum.  A local density approximation (LDA) calculation of jellium
and simple-metal surface
energies~\cite{LK} showed that the xc component $\sigma^{xc}$ can be
several times bigger than the total $\sigma$, and stimulated work~\cite{acfdt}
that led to the development of more sophisticated functionals.

There is now a ladder of nonempirical semilocal density functionals, with each
new rung corresponding to the addition of another ingredient for the energy
density.  The first rung (LDA)~\cite{KS} predicts~\cite{LK,CPT} a positive
$\sigma^{xc}$ for jellium.  The second rung or generalized gradient
approximation (GGA)~\cite{PBE} predicts~\cite{CPT} values about $3\%$ smaller
than LDA, while the third rung or meta-GGA~\cite{TPSS} predicts~\cite{CPT}
values about $2\%$ larger than LDA.  Ascent of the ladder brings steady
improvement~\cite{CPT} in the exactly-known~\cite{PE} exchange part  $\sigma^x$.
The random-phase approximation (RPA), which predicts
inaccurate correlation energies for the uniform gas, predicts values for
$\sigma^{xc}$ about $6\%$ larger than LDA~\cite{PE}.  The semilocal
functionals may be corrected for long-range Coulomb effects~\cite{CPT}, and
the RPA may be corrected semilocally for short-range correlation~\cite{YPK},
producing values $2$ or $3\%$ bigger than LDA. Use, in the framework of TDDFT,
of a uniform-gas-based nonlocal xc kernel~\cite{PP} to correct RPA produces an
RPA-like
$\sigma^{xc}$, because RPA makes compensating errors for density fluctuations
of large and intermediate wavevector~\cite{PP}.

The surface $xc$ energies from all of the above methods disagree strongly with
those from existing high-level correlated methods: At $r_s=4$ (the bulk density
of sodium metal), $\sigma^{xc}$ is about $45\%$ bigger than its LDA value in
the Fermi
Hypernetted Chain (FHNC//0)~\cite{KKQ} and Diffusion Monte Carlo
(DMC)~\cite{AC} calculations for jellium slabs. One conclusion from this might
be that both existing semilocal density functionals (in the framework of DFT)
and uniform-gas-based versions of TDDFT are not valid for predicting
correlation energies, a very dissappointing outcome that would severely limit
the practical usefulness of DFT and TDDFT. In this Letter, we show that this
is not the case, and in the process we resolve the controversy
over the surface energy of simple metals.

Refs.~\cite{PP,Pi} already showed that a careful analysis of the DMC slab
calculations might bring them into agreement with the density functional or
RPA values, and also with surface energies extracted from DMC calculations for
jellium
spheres~\cite{SB,APF}, suggesting that the surface energy puzzle
had
been solved. Nevertheless, one piece of the puzzle remained. Krotscheck and
Kohn~\cite{KK} examined a “collective RPA” and used several xc kernels to
correct for
short-range effects. When they used an isotropic xc kernel derived from the
uniform gas, in the spirit of the TDDFT calculation of Ref.~\cite{PP}
(see also Ref.~\cite{JGDG}), they found surface energies very close to RPA, as
in Ref.~\cite{PP}; when they used an orbital-based Fermi hypernetted chain
approximation (FHNC//0), corresponding to an anisotropic kernel constructed
explicitly for the jellium surface, they found a large positive correction to
the RPA surface energy. Because of this, they concluded that "The local-density
approximation for the particle-hole interaction is inadequate to calculate the
surface energy of the simple metals".

Here we apply a very high-level correlated approach to calculate $\sigma$,
finding values that lie in the narrow range between meta-GGA and RPA, and much
lower than the existing DMC or FHNC//0 slab extrapolations.  We use an
inhomogeneous orbital-based approach (ISTLS)~\cite{DWG} that generalizes the
Singwi-Tosi-Land-Sj\"olander (STLS) formalism~\cite{STLS}.  ISTLS is, like the
RPA it corrects, a “fifth-rung
density functional” that employs the occupied and unoccupied Kohn-Sham
orbitals. A comparison between our calculations, which do not use an isotropic
xc kernel derived from the uniform gas, and the calculations of Ref.~\cite{PP}
leads us to the conclusion that the LDA for the particle-hole interaction is
indeed adequate to describe simple metal surfaces and that existing DFT and
RPA surface-energy calculations are reliable.

\begin{table}
\caption{ The negative of the correlation energy per electron
(in mRyd) of a uniform electron gas in three and two dimensions, where $r_s$ is
the radius respectively of a sphere or circle containing on average one
electron.
FHNC//0: Ref. \cite{KKQ}. STLS 3D: Ref. \cite{DLW}.
STLS 2D: Ref. \cite{Jo}.
DMC 3D: Perdew-Wang parametrization
of Ceperley-Alder Diffusion Monte Carlo (Ref.~\cite{PW}).  
DMC 2D: Parametrization of Eq. (21) of Ref. \cite{KCM}.}
\begin{ruledtabular}
\begin{tabular}{lcccc}
dim. & $r_s$ &FHNC//0 & STLS & DMC\\\hline
3D & 2 & 81.5 & 91.4 & 89.5 \\
   & 3 & 65.3 & 74.7 & 73.9 \\
   & 4 & 55.0 & 64.0 & 63.7 \\
   & 5 & 47.8 & 56.3 & 56.4 \\\\
2D & 1 &      & 211  & 219 \\
   & 2 &      & 155  & 165 \\
   & 4 &      & 108  & 113 \\
   & 8 &      & 66   & 72 \\
\end{tabular}
\end{ruledtabular}
\label{table2}
\end{table}

For the homogeneous electron gas, the STLS approach made a remarkably 
accurate prediction of the correlation energy, as confirmed by later 
DMC calculations (see Table~\ref{table2}). For an arbitrary inhomogeneous
many-electron system, Dobson {\it et
al.}~\cite{DWG} used the linearity and time-invariance of a truncated
Bogoliubov-Born-Green-Kirkwood-Yvon (BBGKY) equation~\cite{BBGKY} to propose
what they called an inhomogeneous STLS (ISTLS) scheme, which can be written as
a Dyson-like "screening" integral equation for the density-response function:
\begin{equation}
\chi(\R,\RR;\omega)=\chi^0(\R,\RR;\omega)+\int d\RP 
Q(\R,\RP;\omega)\chi(\RP,\RR;\omega), 
\label{e3}
\end{equation}
where 
\begin{equation}
Q(\R,\RR;\omega)=-\int d\RP \mbox{\boldmath$\nu$}^0(\R,\RP;\omega)\cdot
g(\RP,\RR)\nabla_{\RP}\frac{1}{|\RP-\RR|}.
\label{e4}
\end{equation}
Here, $\mbox{\boldmath$\nu$}^0(\R,\RR;\omega)$ is a vector response
function defined from the equation
$\chi^0(\R,\RR;\omega)=\nabla_{\RR}\cdot\mbox{\boldmath$\nu$}^0(\R,\RR
\omega)$,
$\chi^0(\R,\RR,\omega)$ being the density-response function of noninteracting
Kohn-Sham (KS) electrons~\cite{note0}, and the equilibrium pair-correlation
function
$g(\R,\RR)$ is obtained from the fluctuation-dissipation theorem as
follows~\cite{acfdt}:
\begin{equation}
g(\R,\RR)=1-\frac{1}{\pi n(\R)n(\RR)}\int^\infty_0du\;\chi(\R,\RR;iu)
-\frac{\delta(\R-\RR)}{n(\RR)}.
\label{e5}
\end{equation}
Equations~(\ref{e3})-(\ref{e5}) are solved selfconsistently, until a converged
solution is obtained~\cite{note1}.

By setting $g(\R,\RR)=1$ (the Hartree limit) in Eq.~(\ref{e4}) and performing
an integration by
parts, the RPA is nicely recovered, as expected. In contrast, the inhomogeneous
FHNC//0 does not recover the actual RPA but a "collective" RPA
instead~\cite{KK}.
We note, however, that the ISTLS scheme cannot be written using an xc kernel
and does not satisfy the reciprocity constraint, i.e.,
$\chi^{ISTLS}(\R,\RR;\omega)\ne\chi^{ISTLS}(\RR,\R;\omega)$, although for
jellium slabs this constraint is found to hold rather well. We also note that
ISTLS is exact for all one-electron densities, for which $g(\R,\RR)=0$, and is
exact in the high-density limit.

For the evaluation of the ISTLS density-response function, we extend the method
described in the Appendix of Ref.~\cite{Egu}. We consider a jellium slab,
assume that $n(z)$ vanishes at a distance $z_0$ from either jellium
edge~\cite{note2}, and expand the single-particle orbitals $\phi_l(z)$ and the
density-response function
$\chi({\bf r},{\bf r}';\omega)$ in sine and double-cosine Fourier
representations, respectively. Because the integral in Eq.~(\ref{e5}) is slowly
convergent and must cancel out a space delta function, we use an expression for
$g(\R,\RR)$ in terms of 
the Hartree-Fock pair-correlation function and the density-response functions
$\chi^0(\R,\RR;\omega)$ and $\chi(\R,\RR;\omega)$. We then use the
adiabatic-connection fluctuation-dissipation formula~\cite{acfdt,PE} to obtain
the xc surface energy from the following equation:
\begin{equation}
\sigma^{xc}=\int_0^\infty d(q/k_F)\,\gamma_q^{xc},
\label{e6}
\end{equation}
where $q$ represents the magnitude of a wave vector parallel to the surface,
and
$\gamma^{xc}_q$ is given by Eqs.~(2) and (3) of Ref.~\cite{PP}, the
density-response function $\chi_{q,\lambda}(z,z';\omega)$ now being the 2D
Fourier transform of our ISTLS density-response function
$\chi_\lambda({\bf r},{\bf r}';\omega)$
of a fictitious jellium slab of fixed density $n(z)$ at coupling strength
$\lambda e^2$.

If one replaces the interacting density-response function
$\chi_{q,\lambda}(z,z';\omega)$ entering Eq.~(3) of Ref.~\cite{PP} by the
noninteracting density-response function $\chi_q^0(z,z';\omega)$, then
the {\it exact} $\sigma^x$ is obtained, as in Ref.~\cite{PE}. Here
we focus our attention on $\sigma^c$, which for comparison
we also calculate (i) in the LDA by replacing $n_{q,\lambda}^c$ in Eq.~(2) of
Ref.~\cite{PP} by the uniform-gas correlation-hole density at the {\it local}
density $n(z)$, and (ii) within TDDFT by introducing in Eq. (3) of
Ref.~\cite{PP} the TDDFT density-response function
$\chi_{q,\lambda}^{TDDFT}(z,z';\omega)$. Within TDDFT, the xc kernel entering
Eq. (4) of Ref.~\cite{PP} is taken to be either zero (RPA) or the uniform-gas
based isotropic xc kernel given by Eqs.~(6) and (7) of Ref.~\cite{PP}.

%
\begin{figure}
\includegraphics[width=\columnwidth]{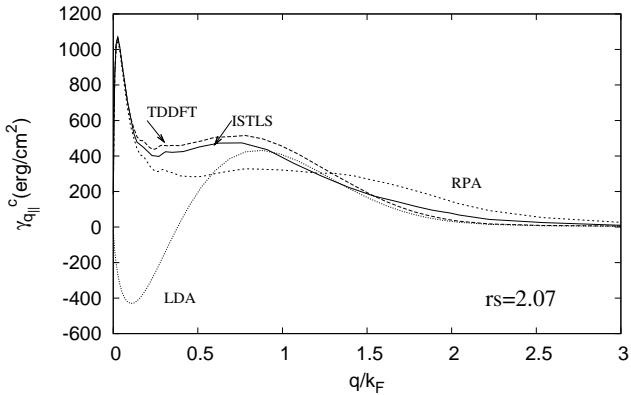}
\caption{Wave-vector analysis $\gamma_q^c$ of the 
correlation surface energy for a jellium slab of thickness 
$7.21r_s$ and $r_s=2.07$. Solid, thick dashed, thin dashed, and dotted lines
represent ISTLS, uniform-gas based TDDFT (as reported in Ref.~\cite{PP}), RPA,
and LDA calculations, respectively. $q$ is the magnitude of the 2D wavevector
(in the surface plane) of the density fluctuations.
The area under each curve amounts to the 
correlation surface energy $\sigma^c$. 
($1 \;\mathrm{hartee}/\mathrm{bohr}^2=1.557\times 10^6 
\;\mathrm{erg}/\mathrm{cm}^2$.)}
\label{f1}
\end{figure}
%

Figure~\ref{f1} shows the wave-vector analysis $\gamma_q^c$ of our ISTLS
correlation surface energy $\sigma^c$ (solid line), together with the
corresponding wave-vector analysis of (i) the LDA correlation surface energy
(dotted line), as obtained by using the Perdew-Wang (PW) parametrization of the
uniform-gas correlation-hole density~\cite{PW2}, (ii) the RPA correlation
surface
energy (thin dashed line), and (iii) the isotropic xc-kernel based TDDFT
correlation
surface energy of Ref.~\cite{PP} (thick dashed line). We observe that
in the long-wavelength limit ($q\to 0$) both ISTLS and TDDFT calculations
coincide with the RPA, which is exact in this limit, while the LDA fails
badly~\cite{note3}. In the large-$q$ limit, both ISTLS and TDDFT calculations
approach the LDA, as expected, while the RPA is wrong~\cite{note4}. The
important lesson that we learn from Fig.~\ref{f1} is that two independent
schemes: (i) our ISTLS approach, which does not use an isotropic kernel derived
from the uniform gas, and (ii) the TDDFT approach of Ref.~\cite{PP},
which uses a uniform-gas based isotropic xc kernel, yield essentially the same
wave-vector analysis of $\sigma^c$. This supports the
conclusion that the local-density approximation for the particle-hole
interaction is indeed adequate to describe simple metal surfaces.

\begin{table}
\caption{LDA, RPA, ISTLS, TDDFT, TPSS~\cite{TPSS}, and recent DMC~\cite{WHFGG}
(per Eq.~(\ref{ee5})) xc surface energies. 
Units are erg/cm$^2$. The numerical grids we use for ISTLS are found to be
inadequate even for RPA when $r_s>3.28$; nevertheless, our best ISTLS estimates for $r_s>3.28$ are found to be very close to the RPA. Values in parentheses represent extrapolations from Eq.~(\ref{ee5}).}
\begin{ruledtabular} 
\begin{tabular}{lcccccc}
$r_s$&$\sigma_{\rm LDA}^{xc}$&$\sigma_{\rm RPA}^{xc}$ &
$\sigma_{\rm ISTLS}^{xc}$ & $\sigma_{\rm TDDFT}^{xc}$ & $\sigma_{\rm
TPSS}^{xc}$ & $\sigma_{\rm DMC}^{xc}$\\\hline
2.00 & 3357 & 3467 & 3417 & 3466 & 3380 & (3392$\pm$ 50)\\
2.07 & 2962 & 3064 & 3026 & 3063 & 2983 & 2993$\pm$ 45\\
2.30 & 2019 & 2098 & 2072 & 2096 & 2034 & 2039$\pm$ 27\\
2.66 & 1188 & 1240 & 1227 & 1239 & 1198 & 1197$\pm$ 13\\
3.00 & 764 & 801 & 800 & 797 & 772 & 768 $\pm$ 10\\
3.28 & 550 & 579 & 580 & 577 & 557 & 551$\pm$ 8\\
4.00 & 262 & 278 & (281)   & 278 & 266 & (261$\pm 8$) \\
6.00 & 53.6 & 58 & (60.5)  & 58 & 55 & (53$\pm ...$)\\
\end{tabular}
\end{ruledtabular}
\label{table1}
\end{table}

To extract the surface energy of a semi-infinite medium, we have considered
three different values of the slab thickness: the threshold width at which the
$n=5$ subband for the $z$ motion is completely occupied and the two widths at
which the $n=5$ and $n=6$ subbands are half occupied, and have followed the
extrapolation procedure of Ref.~\cite{PE}. In Table~\ref{table1}, we show our
extrapolated RPA, TDDFT, and ISTLS xc surface energies, for various values of
$r_s$. Our ISTLS calculations indicate that a persistent cancellation of
short-range xc effects (beyond the RPA) still occurs, as in the case of the
uniform-gas based TDDFT calculations of Ref.~\cite{PP}. However, this
cancellation is found not to be as complete as in Ref.~\cite{PP}, and yields
xc surface energies that are slightly lower than in the RPA but still a little
higher than
in the LDA. Indeed, the difference between our ISTLS surface energies and their
RPA counterparts is very close to the difference between the conventional
GGA~\cite{PBE} surface energies and the corresponding RPA-based GGA surface
energies, thereby supporting the assumption made in Ref.~\cite{YPK} that the
short-range (beyond RPA) part of the correlation energy can be treated within
the GGA. Our ISTLS calculations are also very close to the xc surface energies
obtained by using the non-empirical TPSS meta-GGA xc energy
functional~\cite{TPSS} and a Laplacian-level metal-GGA~\cite{PC}.

The FHNC//0 approach yields a large positive correction to the RPA surface
energy. However, the FHNC//0 approach used in Ref.~\cite{KK} is in fact less
accurate than STLS for the homogeneous 3D electron gas (see Table~\ref{table2}). STLS also does very well for the 2D electron gas (see Table~\ref{table2}). These are
reasons to prefer the ISTLS over FHNC//0 for the surface problems we are
considering. 

The fixed-node DMC calculations reported by Acioli and Ceperley~\cite{AC} have
been critiqued in Refs.~\cite{PP}, ~\cite{Pi} and  ~\cite{WHFGG}. 
Recent DMC calculations by
Wood {\it et al.}~\cite{WHFGG} suggest that the fixed-node approximation
introduces an error that is slightly larger in the slab than in the bulk
calculation and indicate that actual DMC surface energies are larger than in
the LDA but smaller than in the RPA, as occurs with our ISTLS calculations.

The recent DMC calculations~\cite{WHFGG} report total surface energies for
LDA orbitals (from which $\sigma^{xc}$ is easily extracted) and error bars
for $r_s=2.07$, 2.30, 2.66, 3.25, and 3.94.  To refine, interpolate, and
extrapolate these values, we fit to them the physically-motivated form~\cite{APF}
\begin{equation}
\sigma^{xc}(r_s)= A/[r_s^{7/2} (1+Bx+Cx^2+Dx^3)],
\label{ee5}
\end{equation}
where $x = \sqrt{(1+r_s)} - 1$.  We choose typical values $A=50,000$ erg/cm$^2$
(correct $r_s\rightarrow 0$ limit) and $D = 0.248$ (LDA fit), 
then vary
$B$ and $C$ to minimize the sum of the squares of the fit deviation divided
by the DMC error bar, finding $B = 0.6549$ and $C = -0.511$.
Note from Table~\ref{table1} that LDA and TPSS both lie within the error bars of
the recent DMC, while RPA, TDDFT, and ISTLS lie a little higher.
The same fit has been made for ISTLS, with $B =  0.7437$   and $C = -0.653$.

Finally, we note that a detailed analysis of the origin of the xc surface
energy brings us to the conclusion that this quantity is actually "bulk-like",
arising from the moderately-varying-density region inside the classical turning
plane. 
(For $r_s=2$, only $-3\%$ of the total $\sigma^{xc}$ comes from the region
outside. 
  This increases to $-18\%$ for $r_s=4$.)
Inside, the reduced density gradient $s$ falls in a range
($0\leq s < 1.9$) found in the bulk of real solids, where gradient
corrections to LDA exchange and LDA correlation tend to
cancel.

In summary, we have used a very high-level numerically-expensive correlated
approach, the ISTLS method, to analyze the jellium surface energy into
contributions from dynamical density fluctuations of various two-dimensional
wave vectors. This analysis rules out the belief that
the LDA for the particle-hole interaction might be
inadequate to calculate the surface energy of simple metals. Furthermore, our
calculations, which are reasonably close to uniform-gas based TDDFT
calculations~\cite{PP} and not far from the LDA, support the old idea that the
xc surface energy should be well-described within LDA~\cite{acfdt}, and resolve the long-standing surface-energy controversy.

L.A.C., J.M.P., and A.G.-L. acknowledge partial support by the Spanish MEC
and the EC NANOQUANTA. J.P.P. acknowledges NSF support (Grant No. DMR05-01588). We thank H. Le for supplying some data.


\begin{thebibliography}{[Vo]}
\bibitem{note0}
$\chi^0({\bf r},{\bf r}';\omega)$ should be determined from the
exact KS eigenfunctions and eigenvalues. The error introduced by our use of
their LDA counterparts, as obtained by using the PW parametrization~\cite{PW},
is expected to be small.
\bibitem{note1}
Four or five cycles are found to yield a converged solution.
\bibitem{note2} $z_0$ is chosen sufficiently large for the physical results to
be insensitive to the precise value employed.
\bibitem{note3} The LDA is known to be a much better approximation for
$\gamma_q^{xc}$ than for $\gamma_q^c$ alone, even at long wavelengths
($q\to 0$).
\bibitem{note4} In the large-$q$ limit, the RPA (ISTLS) reproduces an
RPA-based (STLS-based) LDA obtained from the RPA (STLS) uniform-gas
correlation-hole density. Differences at large $q$ between the RPA (ISTLS) and
LDA calculations shown in Fig.~\ref{f1} are the result of differences between
the RPA (STLS) uniform-gas correlation-hole density and the more accurate PW
parametrization, which in the case of the RPA are considerably large.
\end{thebibliography}
\end{document}